\documentclass[
    ,final            
  ]
  {aipproc}

\layoutstyle{6x9}

\begin{document}

\title{Analytic Approximations for the Extrapolation of Lattice Data}


\author{Pere Masjuan}{
  address={Faculty of Physics, University of Vienna. Boltzmanngasse 5, A-1090 Wien, Austria.\\
  Departamento de F\'{\i}sica Te\'{o}rica y del Cosmos, Universidad de Granada, Campus de Fuentenueva, E-18071 Granada, Spain. masjuan@ugr.es}
}

%

\begin{abstract}

We present analytic approximations of chiral $SU(3)$ amplitudes for the extrapolation of lattice data to the physical masses and the determination of Next-to-Next-to-Leading-Order low-energy constants. Lattice data for the ratio $F_K/F_{\pi}$ is used to test the approximation proposed\footnote{This work is based on Ref.\cite{paper}}.

\end{abstract}

\maketitle


\section{Introduction}

Quantum Chromodynamics (QCD) is believed to be the correct Quantum Field Theory for the strong interactions in the Standard Model with quark and gluons as degrees of freedom. However, at low energies, due to the strong coupling of QCD, quarks do not show up individually but in bound states called hadrons. In the regime where the light hadron spectrum is observed, a perturbative approach is not feasible. Many efforts have been done to study this sector in the past years. Chiral Perturbation Theory \cite{ChPTSU2,ChPTSU3} as the correct Effective Field Theories of the Standard Model at low energies and simulations of QCD performed numerically on a finite lattice discretizing space and time are the most used ones. In particular, lattice simulations have made huge progress in the light quark sector employing quark masses corresponding to pion masses as low as $150-200$MeV or even physical masses (see, for example, \cite{Lellouch:2009fg}).
In case the simulation does not use physical values for the masses one needs a certain extrapolation to the physical point, extrapolation that is performed nowadays in different ways. The most elaborated once relies on Chiral Perturbation Theory (ChPT) which provides the correct analytic structure of amplitudes in terms of several a priori undetermined constants, the so-called low-energy constants (LECs), independent of the light quark masses by construction.
The number of LECs increases with the order of the expansion in ChPT also increases. It is therefore impossible to extract all of them from experimental data. Lattice calculations offer a new scenario for determining LECs since one can tune the quark masses in a lattice simulation. State-of-the-art lattice studies use next-to-leading-order (NLO) ChPT for chiral extrapolations, determining for example several LECs at this ${\cal O}(p^4)$ order. NNLO ChPT results have only been used recently for the interpretation of lattice data \cite{Bernard:2009ds,Bazavov:2009tw,Noaki:2009sk}.

For comparison with lattice calculations, the explicit dependence on the masses should be known. We propose analytic approximations of chiral $SU(3)$ amplitudes for the extrapolation of lattice data to the physical masses and to determine NLO and NNLO LECs through a more interesting approach than a polynomial.

\section{A dissection of Chiral Perturbation Theory at ${\cal O}(p^6)$}

\begin{figure}
  \includegraphics[height=.25\textheight]{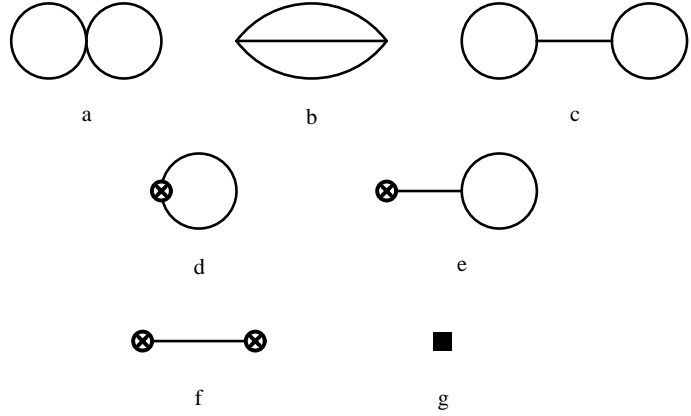}
  \caption{Skeleton diagrams for the generating functional $Z_6$ of ${\cal O}(P^6)$. Simple dots, crossed circles, black box denote vertices from leading-order, NLO, NNLO Lagrangians respectively. Propagators and vertices carry the full tree structure associated with the lowest-order Lagrangian.}\label{skeletonp6}
\end{figure}

The most compact representation of ChPT in the meson sector is in terms of the generating functional of Green functions $Z$ \cite{ChPTSU2,ChPTSU3}. Analogous to the chiral Lagrangian, the generating functional permits a systematic chiral expansion:

\begin{equation}
Z = Z_2 + Z_4 + Z_6 + \dots
\end{equation}

The NNLO functional $Z_6$ of $O(p^6)$ is itself a sum of different contributions shown pictorially in Fig.~\ref{skeletonp6}. In addition to tree diagrams of $O(p^6)$ (diagram g), there are two classes of contributions requiring separate treatments: irreducible (diagrams a,b,d) and reducible (diagrams c,e,f) contributions. For simplicity, we expose here the irreducible case (the reducible one can be found in \cite{paper}). With dimensional regularization, the irreducible diagrams have both double- and single-pole divergences. Moreover, the single-pole divergences of each irreducible diagram are in general non-local. Renormalization theory guarantees, however, that the sum of the three diagrams has only local divergences \cite{Weinberg:1978kz,Z6}, i.e., polynomials in momenta and masses in momentum space. Chiral symmetry guarantees that these divergences can be absorbed by the LECs of $O(p^6)$ via diagram g. In this process an arbitrary renormalization scale $\mu$ is generated. The sum of diagrams a,b,d,g is then finite and can be written in the form
\begin{eqnarray}
Z_6^{\rm a+b+d+g} &=&  \int \!\! d^4x \left\{ \left[C_a^r(\mu) + \displaystyle\frac{1}{4 F_0^2} \left(4\, \Gamma_a^{(1)} \,L  - \Gamma_a^{(2)} \,L^2  + 2\, \Gamma_a^{(L)}(\mu) L \right)\right]  O_a(x)  \right. \nonumber \\[.2cm]
&&   \left.   + \displaystyle\frac{1}{(4\pi)^2} \left[L_i^r(\mu) -   \displaystyle\frac{\Gamma_i }{2} L \right] H_i(x;M) +  \displaystyle\frac{1}{(4 \pi)^4}  K(x;M) \right\}~.
\label{eq:Z6irr}
\end{eqnarray}

The monomials $O_a(x) ~(a=1,\dots,94)$ define the chiral Lagrangian of $O(p^6)$
\cite{Bijnens:1999sh} with associated renormalized LECs $C_a^r(\mu)$, the $L_i^r(\mu) ~(i=1,\dots,10)$ are renormalized LECs of $O(p^4)$ with associated beta functions $\Gamma_i$ \cite{ChPTSU3} and the coefficients $ \Gamma_a^{(1)}$, $\Gamma_a^{(2)}$ and $\Gamma_a^{(L)}$ are listed in Ref.~\cite{Z6}. $F_0$ is the meson decay constant in the chiral $SU(3)$ limit. The chiral log

\begin{equation}
L = \displaystyle\frac{1}{(4\pi)^2} \ln{M^2/\mu^2}
\label{eq:clog}
\end{equation}

involves an additional (arbitrary) scale $M$ but $Z_6^{\rm a+b+d+g}$ as well as the total generating functional $Z_6$ are independent of both $\mu$ and $M$. $H_i(x;M)$ are one-loop functionals associated with diagram d whereas the two-loop contributions (except for the chiral logs) are contained in the functional $K(x;M)$. The functional (\ref{eq:Z6irr}) is scale independent \cite{paper}.

The complete generating functional of $O(p^6)$ is then given by the sum of the irreducible and reducible parts, again independent of both scales $\mu$ and $M$:
\begin{equation}
Z_6 = Z_6^{\rm a+b+d+g} + Z_6^{\rm c+e+f}~.
\label{eq:Ztotal}
\end{equation}

\subsection{Analytic Approximation for chiral $SU(3)$}

The genuine two-loop contributions contained in the functional $K(x;M)$ are usually only available in
numerical form for chiral $SU(3)$. On the other hand, the one-loop contributions can be given in analytic form and the dependence on meson masses is manifest. For the chiral extrapolation of lattice results, we therefore suggest the following approximate form for the functional of $O(p^6)$:
\begin{eqnarray}
Z_6^{\rm app} &=&  \int \!\! d^4x \left\{ \left[C_a^r(\mu) +
\displaystyle\frac{1}{4 F_0^2} \left(4\, \Gamma_a^{(1)} \,L  -
\Gamma_a^{(2)} \,L^2 + 2\, \Gamma_a^{(L)}(\mu) L
\right)\right] O_a(x) \right. \nonumber \\[.2cm]
&& + \left. \displaystyle\frac{1}{(4\pi)^2} \left[L_i^r(\mu) -
  \displaystyle\frac{\Gamma_i }{2} L \right] H_i(x;M) \right\} \nonumber
\\[.2cm]
&+&  \int \!\! d^4x\,d^4y \left\{\left(L_i^r(\mu) -
  \displaystyle\frac{\Gamma_i }{2} L \right)
  P_{i,\alpha}(x) \,G_{\alpha,\beta}(x,y)
  \left(L_j^r(\mu) -
  \displaystyle\frac{\Gamma_j }{2} L \right) P_{j,\beta}(y)\right.
  \nonumber \\[.1cm]
&& + \left. 2\,\left(L_i^r(\mu) -
  \displaystyle\frac{\Gamma_i }{2} L \right)
  P_{i,\alpha}(x) \,G_{\alpha,\beta}(x,y)\, F_\beta(y;M)\right\} ~.
\label{eq:logapp}
\end{eqnarray}

The approximation proposed, called Approximation I from now on, consists on dropping the $K(x;M)$ pieces and the reducible 2-loop contributions from diagrams a and c in Fig.\ref{skeletonp6} while keeping all the chiral logarithms $L$ (more details can be found in Ref.~\cite{paper}). For a reliable determination of renormalized LECs the analytic approximation (\ref{eq:logapp}) should be scale independent as it is the case. The main issue to address now is the applicability of that approximation. The simple answer is based in a naive chiral counting. Considering a $SU(3)$ amplitude normalized to 1 at LO, the chiral counting suggest that the NLO to be of the order of $0.3$, the NNLO of the order $0.3^2=0.09$ and the NNNLO of the order $0.3^3=0.027\simeq 3\%$. That counting suggests that a good approximation to the NNLO amplitude should have an accuracy not worse than $3\%$ which would correspond to the higher order not considered in the approximation. This criterion will be applied for the $F_K/F_{\pi}$ quantity in the next subsection.

\subsubsection{Approximation II}

The above approximation is motivated by large $N_c$ behavior but some of the terms not included in the approximate functional (\ref{eq:logapp}) have a known analytical form. In practice, one may include those terms, i.e., the products of one-loops amplitudes from diagrams a and c in Fig. \ref{skeletonp6}, to improve the accuracy of the approximation for certain observables. This second approach (called Approximation II) should also satisfy the accuracy criteria.

\subsection{Application to lattice data for $F_K/F_{\pi}$}

We now apply the analytic approximations Approximation I and Approximation II to the ratio $F_K/F_{\pi}$ of meson decay constants. $F_K/F_{\pi}$ is a suited quantity for our exploratory exercise for several reasons: at $\mu=0.77$GeV the genuine two-loop accounts less than $1\%$  and it is known numerically (\cite{Amoros:1999dp,Bernard:2009ds}) allowing a direct evaluation of the approximation; the BMW Collaboration \cite{Durr:2010hr} provides 13 lattice points (considering only $m_{\pi}<450$MeV) for that ratio.

The full NNLO calculation at physical meson masses tells us that $F_K/F_{\pi}-1=0.14$ at ${\cal O}(p^4)$. At ${\cal O}(p^6)$ there are three different contributions: tree level, 1-loop$\times L_i$, 2-loops and contribute to $F_K/F_{\pi}-1$ with $0.008$, $0.051$ and $0.002$ respectively.

In our approximation, the ${\cal O}(p^4)$ and the tree level and 1-loop$\times L_i$ at ${\cal O}(p^6)$ are exact by construction. Our estimation for the genuine 2-loops piece amounts, using $M=M_K$, $-0.030$ using Approximation I and $-0.011$ using the Approximation II. In both cases, the criteria of applicability is fulfilled.

We can now proceed on fitting the 13 data points with our approximations for $F_K/F_{\pi}$. We obtain for the LECs at $\mu=0.77$ GeV (the fitting function can be found in Ref.\cite{paper}) the results shown in Table \ref{tableresults}.
\begin{table}
\centering
\begin{tabular}{|c|c|c|c|}
  \hline
    & $L_5 \cdot 10^3$ & $(C_{14}^r + C_{15}^r)\cdot 10^{3} ~{\rm GeV}^{2}$ & $(C_{15}^r + 2\,C_{17}^r)\cdot 10^{3} ~{\rm GeV}^{2}$ \\
    \hline
  Approx I & $0.76 \pm 0.09$ & $0.37 \pm 0.08$ & $1.29 \pm 0.16$ \\
  Approx II &$0.75 \pm 0.09 $&$ 0.20 \pm 0.07 $&$ 0.71 \pm 0.15 $\\
  \hline
\end{tabular}
\caption{Summary of results for the fitted $F_K/F_{\pi}$ for both Approximation I and Approximation II.}\label{tableresults}
\end{table}

The three parameters are strongly correlated \cite{paper}. Taking that into account, $F_K/F_{\pi}$ for physical meson masses is found to be
\begin{eqnarray}
F_K/F_{\pi}|_{Approx I} &=& 1.198 \pm 0.005 ~,\\ \nonumber
F_K/F_{\pi}|_{Approx II} &=& 1.200 \pm 0.005 ~,
\end{eqnarray}

in good agreement with the result $F_K/F_{\pi}=1.192(7)_{stat}(6)_{syst}$ of Ref. \cite{Durr:2010hr}. Our errors take only into account the statistical error of the lattice values for $F_K/F_{\pi}$ but not the systematical once since we do not pretend to improve on the accuracy of the result but show the feasibility of the lattice data to explore the ${\cal O}(p^6)$ chiral lagrangian.

We have include the NLO LEC $L_5$ in our fit because it is the only LEC that appears at this order and also because at NNLO, $L_5^2$ is the leading piece in $1/N_c$ counting. We use the common fit10 \cite{fit10} values for the other $L_i$ appearing at NNLO.

The residual dependence on the scale $M$ is evaluated by setting $M=M_K$ (the lattice value) and varying this scale by $\pm 20\%$. In doing this, both $F_K/F_{\pi}$ and $L_5$ remain again unchanged while the LECs of ${\cal O}(p^6)$ vary within two standard deviations.
\vspace{-0.25cm}
\subsubsection{Other examples}

The approximations proposed here have also been applied\footnote{Work in progress in collaboration with G.Ecker and H.Neufeld.} to two more quantities.

The first one is the ratio $F_{\pi}/F_0$ in chiral $SU(3)$ to obtain a value for the LEC $L_4$ and for the chiral parameter $F_0$, both badly known. The comparison of our approximations in that case with the numerical available results, \cite{Bernard:2009ds} are encouraging.

The second example is the $K_{l3}$ vector form factor $f_+(t)$. The particular case $f_+(0)$ is of interest to obtain a prediction for the CKM parameter $V_{us}$. The approximations discussed here do not appear very promising in this case: the chiral expansion shows an unusual behavior and the approximations do not match with the available numerical results of Refs. \cite{Bernard:2009ds,Post:2001si,Bijnens:2003uy}.

\vspace{-0.25cm}

\section{Conclusions}

We have proposed analytic approximations for chiral $SU(3)$ amplitudes starting from the structure of the generating functional of Green functions to ${\cal O}(^6)$ in such a way that only tree-level and one-loop diagrams are needed. Our proposal provides useful formulas for the extrapolation of lattice data and also allows, due to their renormalization scale independence, for the determination of the LECs at NNLO otherwise difficult to extract from experimental data. These approximations are superior to the NLO amplitudes and to the well-known double-log approximations \cite{Bijnens:1998yu}. Concerning the $1/N_c$ expansion, these amplitudes contain all the leading and next-to-leading terms and all the chiral logarithms.
This approach is specially useful in cases where the genuine two-loop contributions are small, in accordance to the large-$N_c$ counting. The particular case of the quantity $F_K/F_{\pi}$ has this property. We fit the approximated expression of $F_K/F_{\pi}$ to recent lattice data where we obtain a value for $F_K/F_{\pi}$ in agreement with the analysis of Ref. \cite{Durr:2010hr}. We also obtain a value for $L_5$ and for the LECs of ${\cal O}(p^6)$, $C_{14}^r + C_{15}^r$ and $C_{15}^r + 2C_{17}^r$. While both $F_K/F_{\pi}$ and $L_5$ are insensitive to the approximation made, the LECs of ${\cal O}(p^6)$ are consistent with expectations but subject to uncertainties exceeding the lattice errors. All in all that suggests the potentiality to explore higher-orders on the chiral expansion using lattice data with an approximation that provides with more insight than for example a simple polynomial.

We also suggest an Approximation II, as a modification of Approximation I, that includes the reducible diagrams a and c (Fig~\ref{skeletonp6}), when the strict Large-$N_c$ is ignored.

\vspace{-0.25cm}

\begin{theacknowledgments}
I would like to thank the organizers for the nice atmosphere during the conference and G.Ecker and H. Neufeld for comments on the manuscript. This work has been supported by the EU contract MRTN-CT-2006-035482 (FLAVIAnet), by MICINN, Spain (FPA2006-05294), the Spanish Consolider-Ingenio 2010 Programme CPAN (CSD2007-00042) and by Junta de Andaluc\'{\i}a (Grants P07-FQM 03048 and P08-FQM 101).
\end{theacknowledgments}



\bibliographystyle{aipproc}   


\IfFileExists{\jobname.bbl}{}
 {\typeout{}
  \typeout{******************************************}
  \typeout{** Please run "bibtex \jobname" to optain}
  \typeout{** the bibliography and then re-run LaTeX}
  \typeout{** twice to fix the references!}
  \typeout{******************************************}
  \typeout{}
 }


\end{document}